# Far-field directionality control of coupled InP nanowire lasers

*Lukas R. Jäger[1,2], Wei Wen Wong[2], Carsten Ronning[1], Hark Hoe Tan[2,*]*

[1] Institute of Solid State Physics, Friedrich Schiller University Jena, Max-Wien-Platz 1, 07743 Jena, Germany

[2] ARC Centre of Excellence for Transformative Meta-Optical Systems, Department of Electronic Materials Engineering, Research School of Physics, The Australian National University, Canberra, ACT 2600, Australia

* Corresponding author

## Abstract

Nanowire (NW) lasers hold great promise as compact, coherent on-chip light sources that are crucial for next-generation optical communication and imaging technologies. However, controlling their emission directionality has been hindered by the complexities of lasing mode engineering and fabrication. Here, we demonstrate spatially-engineered far-field emission from vertically emitting InP NW lasers by establishing precise control over the optical coupling between site-selective NWs, without relying on post-epitaxy transfer and alignment processes. Leveraging this process capability, we design and grow NW pairs and triplets that lase in the TE01 waveguide mode. We then demonstrate the ability to modify their far-field emission profiles from the signature doughnut-like emission to a double-lobed emission profile by changing their optical coupling gap, evidenced by closely matching simulation and experimental profiles. Moreover, through numerical simulations, we show further enhancement in the far-field directionality by arranging the NW laser pairs in a periodic array, demonstrating the feasibility of a directional lasing metasurface. Our results provide a foundation for efficient integration of coherent light generation and beam steering in on-chip light sources.

# Introduction

The development of nanophotonic devices has garnered significant attention due to their potential to revolutionize optical communication, sensing, and imaging technologies [1, 2]. Semiconductor nanowires (NWs), in particular, have emerged as promising building blocks for miniature photonic devices, including nanoscale lasers, owing to their unique optical and electronic properties [3, 4, 5, 6].

Early demonstrations of NW lasers focused on single, mechanically transferred, horizontally lying NWs on various substrates. Laser emission from the two end-facets of the NWs, which is typically collected from the top, is irregularly scattered by the substrate and may interfere uncontrollably, resulting in poor emission directionality [7]. In order to improve the vertical directionality, Xu et al. transferred the NWs onto a bull's eye structure [8]. This process, however, relies on an unscalable pick-and-place method with stringent alignment requirements. More recent studies have investigated lasing in vertical, on-chip NWs fabricated via site-selective area epitaxy (SAE), which overcome scalability limitations of the mechanical pick-and-place process. Zhang et al. demonstrated vertical nanowire laser arrays with embedded quantum wells, though control over the emission directionality was not demonstrated [9]. One approach to achieve directional lasing was demonstrated in reference [10], where an InP micro ring laser was coupled to a vertical NW, epitaxially grown in the ring's center, yielding a vertically emitted, highly directional beam. Thus, site-selective epitaxial growth is particularly attractive for such advanced nanoscale optoelectronic devices for directional and controllable lasing emission characteristics. Moreover, this growth process results in the formation of high-quality side facets, which inherently reduce optical losses. In addition, the bottom-up growth approach enables the integration of high-quality III-V materials on unconventional substrates, such as silicon [11]. Catalyst-free SAE further expands the versatility of this technique by permitting the arrangement of NWs into arbitrary patterns, effectively acting as nano-antenna elements within a metasurface-like lattice.

We build on the advantages of those approaches, namely utilizing a scalable and highly deterministic SAE process without the need for complex transfer or alignment processes. In this study, we demonstrate that optical coupling of vertically emitting lasing InP NWs can be employed to engineer the far-field emission, marking a significant step toward the development of lasing metasurfaces. Our approach facilitates the construction of a light-emitting metasurface, where each meta-atom, comprising two or three NWs, yields coherent emission at large angles.

# Results

## Design of InP NW pair lasers

We first designed a model system of an InP NW pair, schematically illustrated in Figure 1a. The single-crystalline InP substrate had a 200 nm-thick layer of silicon oxide with two circular openings of 100 nm in diameter positioned side by side, which were defined by electron beam lithography (EBL) and served as a mask for SAE of vertically standing NWs. The NWs grew both vertically and laterally, as shown in Figure 1a (ii). The lateral overgrowth of the InP NWs on the $SiO_2$ layer introduces partial reflectivity to the waveguide lasing modes, thereby enabling the upstanding NW to act as a Fabry–Pérot cavity [12]. A representative NW pair can be seen in the scanning electron microscopy (SEM) image in Figure 1a (iii). To guide the design, we first analyzed this model system using finite-difference time-domain (FDTD) simulations and an eigenmode expansion solver. These tools allowed us to determine two key parameters: the necessary NW diameter and the optimal inter-NW gap for efficient near-field coupling and far-field shaping.

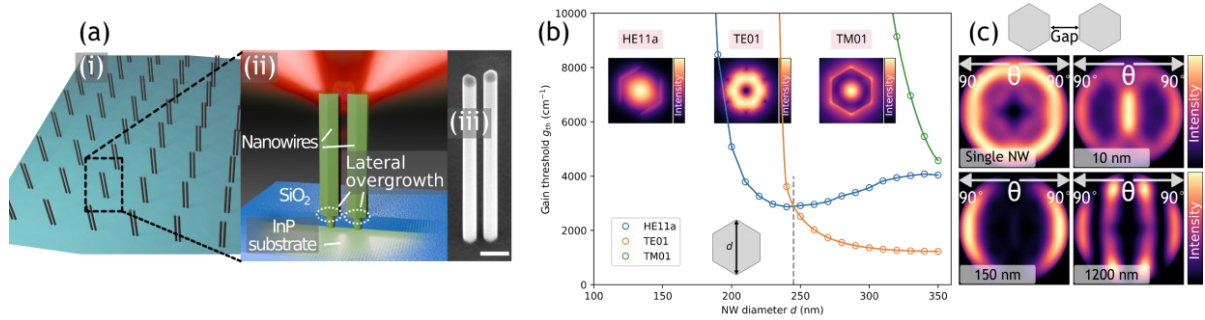

**Figure 1:** Design of a lasing InP NW pair. **(a)**(i) Schematic view of InP NW pairs grown vertically on an InP substrate. (ii) Schematic showing the pair of NWs setup that were grown through a SiO$_2$ mask with some lateral overgrowth., resulting in a cavity, and the simulated far-field emission pattern of such a pair at the lasing wavelength of 850 nm. (iii) SEM image of a pair of NWs. The scale bar is 500 nm. **(b)** Simulated gain threshold g$_{th}$ of a single InP NW with a hexagonal cross section as a function of its diameter for the first four waveguide modes (the fundamental modes, HE11a and HE11b, are degenerate). Insets are simulated mode-profiles. The dashed grey line at ca. 245 nm indicates the NW diameter where the TE01 mode starts to be the lowest gain threshold mode. **(c)** Simulated far-field emission patterns for a single InP NW and for three different NW pairs with varying gaps, as indicated. The modes are TE01 in the single NW case and TE01-a in the NW pairs.

Figure 1b plots the simulated gain threshold g$_{th}$ of a single InP NW with a hexagonal cross section for the fundamental HE11 (hybrid mode), TE01 (transverse electric), and TM01 (transverse magnetic) modes, as a function of its diameter. The formula used for the calculation of the gain is described in the Supplementary Information (SI) Section I. The insets show the simulated mode-profiles of the fundamental mode (HE11a is shown, the HE11b mode looks the same, but rotated), which are degenerate in a hexagonal waveguide. The transverse TE01 mode is also shown. In this context, two modes are degenerate if they have the same effective index. A hexagonal cross section was chosen based on the general morphology of wurtzite-phase InP nanowires which features six $\{1\bar{1}00\}$ crystal sidewall facets. However, in practice the cross section of the grown NWs can deviate slightly from this profile (see Figure 1a (iii)), but the mode profile does not significantly change due to this deviation (see Figure S1). Above a diameter of about 250 nm, the dominant lasing mode is the doughnut shaped TE01 mode, which has the lowest gain threshold, making it suitable for lasing operation. The low gain-threshold of the TE01 mode can be mainly attributed to a high end-facet reflectivity at the top end facet and sufficient reflectivity at the bottom end facet between the NW and silicon oxide mask, as a result of the lateral overgrowth. The fundamental modes do not have a cutoff diameter, though in practice, the strong delocalization of modes at smaller diameters raises the gain-threshold beyond usable limits. They generally have high gain threshold across the simulated diameters due to the electric field being concentrated in the center of the NW. This leads to low reflectivity at the NW substrate interface, as a result of leakage into the InP substrate. The TM01 mode (green line in Figure 1b) exhibits a significantly larger gain threshold in the investigated diameter range and a larger diameter cutoff. During lasing, the different modes compete for gain, though even small differences in gain threshold can cause one mode to dominate, suppressing the others. We designed the NW to lase in the TE01 mode due to its low gain threshold and relatively small cutoff diameter. Growing a relatively small diameter wurtzite (WZ)-phase InP NW is more feasible compared to the diameter required for TM01-mode lasing (> 350 nm).

After setting the NW diameter to support single transverse mode TE01 lasing, we optimized the optical coupling between two identical NWs to engineer their far-field emission directionality. When two lasing NWs are optically coupled, the TE01 mode splits into two distinct modes, TE01-a and TE01-b, due to the lifting of polarization degeneracy. In this study, we designate the mode with the higher effective index ($n_{eff}$) as TE01-a. This mode typically exhibits a lower gain threshold due to its stronger optical confinement. The reason for the mode splitting, as mentioned above, is that the polarization degeneracy is lifted, as shown in Figure S2. Consequently, the electric fields in the two NWs are either in or out of phase. The origin of the two distinct far-field patterns of the TE01-a and TE01-b modes can thus be explained by approximating the emission of the top end facets of the NW pair by two electric dipoles which are out of phase and in phase, respectively. To corroborate this, we simulated the far-field emission patterns from this subwavelength spaced dipole system (Figure S3).

Figure 1c shows the simulated far-field emission patterns for a single InP NW and for three NW pairs with different gaps. A single NW emits in a ring-like pattern almost equally in all azimuthal directions. However, a NW pair, lasing in the TE01-a mode and with a gap of approximately 150 nm, produces a double-lobed directional emission pattern, where most of the light is emitted at two symmetric large-angle lobes and almost none in the normal direction. For smaller gaps, where the near-field coupling is expected to be strongest, the far-field pattern is more spread-out with a stronger normal-direction lobe, compared to the 150 nm gap case. When the NW gap exceeds the lasing wavelength (~850 nm), interference effects emerge, due to the exponential-like decay of the mode's evanescent field with increasing distance, which significantly reduces coupling. The NWs therefore act effectively as separate point-like coherent scatterers (Figure 1c, 1200 nm gap). Based on these findings, we fabricated InP NW pairs with diameters of approximately 250 nm and gaps ranging from 100 to 150 nm.

## Realizing directional emission in NW pairs

Figures 2a and 2b show the photoluminescence (PL) spectra of an InP NW pair at various pump fluences. At low excitation fluence, the near-band-edge (NBE) emission is broad (see also Figure S4). As the excitation fluence increases, the PL peak undergoes a blueshift due to bandgap renormalization and band-filling effects [13]. Furthermore, equally-spaced Fabry–Pérot modes begin to appear in the emission spectra. Between approximately 200 and 500 µJ cm$^{-2}$ per excitation pulse, the NW laser pair operates in the amplified spontaneous emission (ASE) regime, an intermediate state between spontaneous emission and lasing. At a threshold fluence of around 350 µJ cm$^{-2}$ pulse$^{-1}$, the lasing modes rapidly increase in intensity, as evident in Figure 2b by the transition from the broad PL emission to mode-dominated emission. At even higher excitation levels, a slight additional blue shift occurs, along the emergence of additional Fabry–Pérot lasing modes.

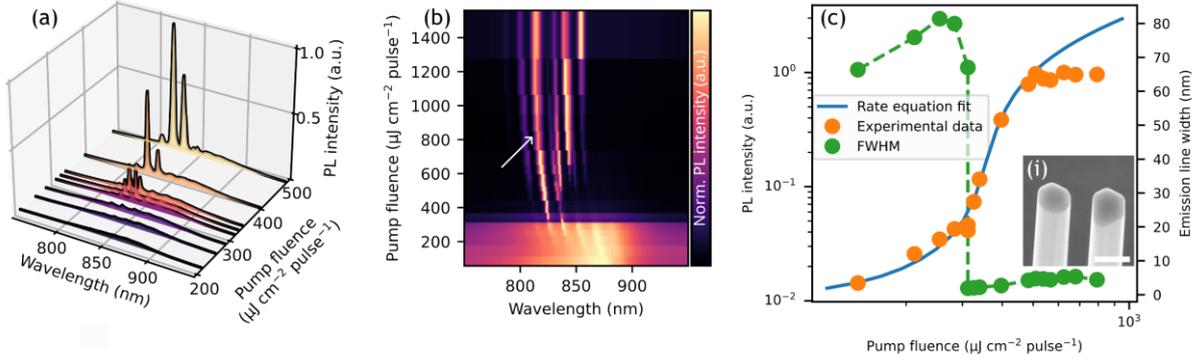

**Figure 2:** Photoluminescence (PL) spectra of an InP NW pair. **(a)** PL spectra as a function of pump fluence. **(b)** PL spectra with normalized intensity showing a clear and abrupt transition from spontaneous emission to lasing emission characterized by Fabry-Perot modes. The white arrow denotes the mode used for the calculation of the LL-curve shown in (c). **(c)** PL intensity of the most prominent Fabry-Perot mode over the laser pump fluence (left axis), fitted by solving laser-rate equations (see SI). The fit yielded a gain threshold $g_{th}$ = 3850 cm$^{-1}$, $\beta$ = 0.009. Additionally, the emission line width (right axis), shows a step function like behavior indicative of a lasing transition. Inset (i) is an SEM image of the measured NW pair. The scale bar is 200 nm.

The light-in versus light-out (LL) curve of the dominant lasing mode of the InP NW pair, shown in Figure 2c, exhibits the characteristic S-shape indicative of lasing. We solved the laser rate equations (see SI Section VI) to fit the LL curve and obtain the spontaneous emission factor ($\beta$-factor) and gain threshold. The fit was restricted to the single-mode lasing regime (up to 500 µJ cm$^{-2}$ pulse$^{-1}$), as the laser rate equations are only applicable for single mode lasing. This accounts for the divergence between the experimental data and the obtained fit beyond 500 µJ cm$^{-2}$ pulse$^{-1}$ in Figure 2c. The fitted gain threshold $g_{th} \approx 3850\ cm^{-1}$, is slightly higher than the simulated cutoff threshold for the single NW TE01 mode ($g_{th} \approx 3000\ cm^{-1}$), likely due to the discrepancies between the ideal simulation and the actual NW morphology. According to the simulation data in Figure 1b, increasing the NW diameter from 250 to 300 nm would likely halve this lasing threshold.

As shown in Figure 3, the measured emission patterns closely match simulations and exhibit the clear directionality intended by the design, despite deviations from the ideal hexagonal NW cross section (SEM images in Figures 3a (ii) and 3b (ii)). The emission angle $\theta$ is indicated in the measured patterns in Figure 3. The maximum collection angle, determined by the numerical aperture (NA) of the objective, is $\theta_{max} \approx 64°$. Note that the emission is plotted on a linear scale with respect to $k_x/k_0$ and $k_y/k_0$. Figures 3a and 3b show two NW pairs with a gap of approximately 130 and 210 nm, respectively. The slight imbalance of the emission intensity between the left and right lobe of the patterns can be attributed to an asymmetry in NW diameters. Simulations of the mode profiles indicate that in this case the larger-diameter NW laser contributes more power than the smaller one. This asymmetry is more pronounced in Figure 3b, where the diameter of the right NW has a substantially larger diameter.

Interference fringes are visible in both measured patterns (Figures 3a and 3b). These arise from interference between light scattered from the top and bottom end-facets. We corroborated this through simulations (see Figure S8) and by comparing the spacing between interference rings, which corresponds to the length of the NW system (see Figure S9).

The top-right insets of both Figures 3a and 3b show the simulated emission patterns based on the actual geometry, taking into account the facet asymmetries in the nanowire pair. They show a slightly stronger emission asymmetry than observed experimentally. Comparing the far-field emission patterns of the NW pairs with 130 and 210 nm gaps, we find that the pair with the larger gap exhibits weaker directionality. In the ideal simulations (compare Figure 1c for 150 nm gap and Figure S10 for 200 nm gap), which assume perfect hexagonal cross sections and identical NW dimensions, NW pairs with 130 nm 210 nm gaps should produce nearly identical far-field patterns. Additionally, the facet-accurate simulations produced a relatively small difference in their respective far-field patterns as well. This suggests that certain fabrication-induced imperfections in the real NW geometry may not fully be captured by the simulations, because the NW pair with the 210 nm gap (Figure 3b) exhibited noticeably weaker directionality than predicted.

The TE01-b mode generally exhibits a higher gain threshold, owing to its lower effective index. Due to growth imperfections, some of the NW pairs exhibit lasing in this mode, which also emits directionally into the far-field, with lobes oriented orthogonally to those of the TE01-a mode (see Figure S11).

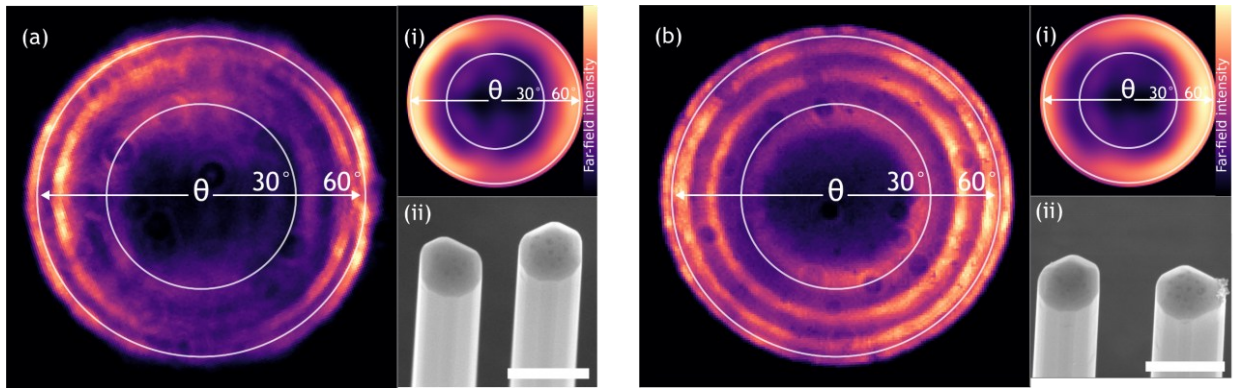

**Figure 3:** Comparison between measured back-focal plane, where the fields are directly proportional to the far-fields field, and simulated far-field patterns for two InP NW pairs. The larger left images are experimental data, the top right images are simulated patterns for the actual NW geometry, extracted from the respective SEM images (inset (ii), scalebars are 300 nm). The ripple pattern in the experimental data is caused by interference between the facet ends of the NW, which is an effect excluded from our simulations. **(a)** The NW pair has a gap of about 130 nm and shows a clear directionality in the emission pattern. **(b)** The right NW pair has a gap of about 210 nm and shows weaker emission directionality.

## Modelling directionality enhancement in an array of NW pairs

So far, we demonstrated directional emission by utilizing optical coupling in a pair of NWs. The produced far-field showed the expected emission lobes, which, however, are markedly broad. Here, we consider arranging NW pairs into a periodic array to leverage inter-pair interference. We investigated how the array structure depicted in Figure 4a influences the emission properties of the NW pairs. In particular, we examined how the pitch between NW pair unit cells and the number of contributing cells affect the far-field emission. The unit cell consisted of a pair of identical NWs with a diameter of 280 nm and a gap of 100 nm.

First, we fixed the contributing number of NW pairs to 3×3 period, varied the pitch, and calculated the resulting far-field patterns, as shown in Figure 4b. We applied Gaussian illumination weighting to the far-field contributions of the unit cells, using Lumerical's built-in far-field calculation functionality, to reduce the aperture effect that would result from a sharp (step-function) cutoff. We found that a pitch of 1000 nm yields a far-field pattern with lobes similar in polar angle, but significantly narrower in the azimuthal direction, compared to a single NW pair.

Larger pitches result in far-field patterns with more than two lobes. Reducing the pitch below 1000 nm altered the polar emission angle and led to the formation of a new central lobe, emitting in the normal direction.

The profiles of the injected mode (see Figure S12a) closely resemble the TE01-a mode throughout reducing the pitch from 1500 to 800 nm. To elucidate the mechanism behind the pitch-sensitive far-field profiles, we conducted analogous simulations with dipole source pairs. The produced far-fields (Figure S12b) resemble those of the simulated NW pair array. This suggests that the evolution of the far-field pattern when reducing the pitch, including the central lobe formed at a pitch of 800 nm, is caused by interference, rather than a kind of photonic-crystal-like mode.

We then fixed the pitch at 1000 nm and increased the number of NW pairs contributing to the far-field emission. This simulates how the far-field is expected to evolve as the excited array size increases. As shown in Figure 4c, increasing the number of NW pairs leads to narrower emission lobes, while the main emission solid angle remains unchanged.

The growth of the arrays yielded NWs with highly uniform lengths and diameters, as shown in Figure 4d. However, the NW diameter was approximately 100 nm, which is well below the TE01 mode cutoff. Nonetheless, PL measurements were performed (see SI Section X). Due to a significant mismatch between the fabricated NW diameter and the designed TE01 mode cutoff, lasing was not realized. Only broad PL far-field emission was observed. Future samples will be grown with greater emphasis on optimizing the lateral overgrowth. A homogenous array of NWs with clean crystal facets and diameters of ~260 – 280 nm could enable the realization of a directional lasing metasurface with a high emission directionality.

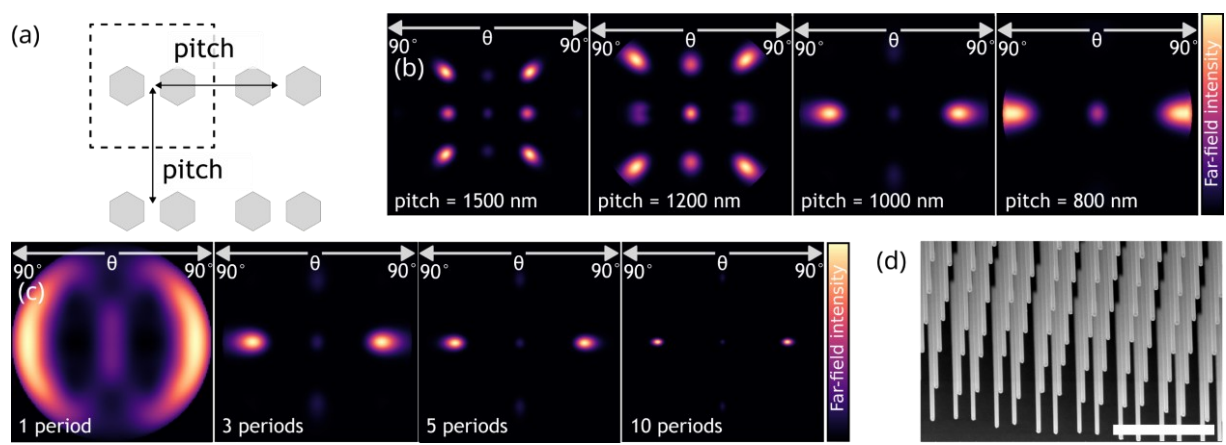

**Figure 4:** Transitioning from a single InP NW pair to an array of pairs. **(a)** Schematic top-down view of the array structure. **(b)** Far-field emission patterns from a 3x3 unit-cell for different pitch size. **(c)** Far-field emission patterns of an increasing number of contributing NWs at a pitch of 1000 nm, displaying increasingly narrow emission directionality. The NW diameters were 280 nm, and the gaps were 100 nm. **(d)** SEM image of a grown array. The scale bar is 2 μm.

## Realizing directional emission in a triplet NW structure

To demonstrate the versatility of our concept, we fabricated NW triplets and measured their far-field lasing emission. Figure 5a illustrates the arrangement of the three NWs in the triplet configuration. In this configuration, the TE01 mode splits into three: TE01-1, TE01-2, and TE01-3. These labels indicate the number of NWs that contain significant portions of the mode's electric field (shown in Figure S12). Among the three, the TE01-2 mode features the highest effective index, followed closely by the TE01-1 mode. Figures 5b to 5d show three NW triplets, each emitting a far-field corresponding to one of the three modes. The simulated emission patterns (the top right insets) are based on NWs with a 280 nm diameter and 120 nm gap with ideal geometry. The larger diameter was selected because the TE01-3 mode has a higher cutoff diameter than the single-NW TE01 mode. This mode also best matches the far-field emission observed from the NW triplet depicted in Figure 5d (ii). Notably, the triplet emitting the highest effective index mode (Figure 5b (ii)) more closely resembles the ideal hexagonal geometry compared to the other two systems (Figures 5c (ii) and 5d (ii)). This suggests that deviations from the ideal geometry likely led to the lasing of the system in lower-index modes.

The far-field emission pattern of the TE01-2 mode closely resembles that of the TE01-a mode of the NW pair. A similar correspondence exists between the TE01-1 and TE01-b modes. Thus, NW triplets are not strictly required to reproduce these far-field patterns, since rotated NW pairs can produce nearly identical emission profiles. The pattern of the TE01-3 mode cannot be reproduced by a NW pair. However, its low effective index also makes it the most difficult mode to engineer. Furthermore, the TE01-3 mode lacks the azimuthal directionality of the TE01-1 and TE01-2 modes. Because the NW triplet configuration affords more degrees of freedom in its design, mode dominance can be engineered by intentionally introducing certain asymmetries into the system, for example by designing the SAE mask openings with unequal gaps or diameters.

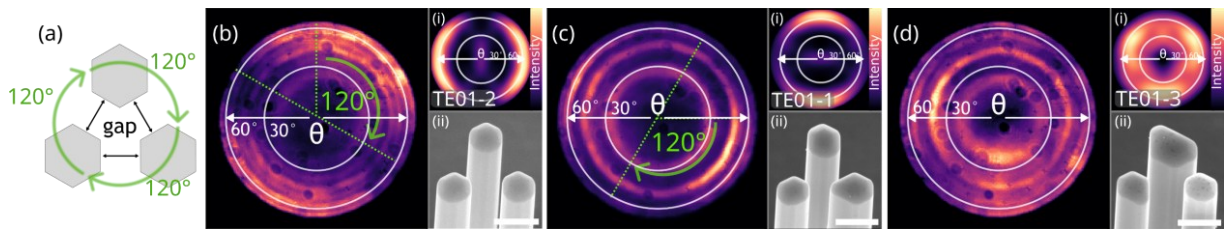

**Figure 5:** Far-field measurements of the NW triplet structure. **(a)** Schematic illustration of the arrangement of NWs in the triplet structure. The system has a three-fold rotational symmetry. **(b)** Far-field emission pattern (large), (b)(i) simulated far-field pattern of the TE01-1 mode, and (b)(ii) SEM image of the NW triplet (bottom-right). **(c)** Similar to (b), but for a different NW triplet, and different underlying mode (TE01-2). **(d)** Similar to (b), for a different NW triplet, with TE01-3 as the underlying mode. No rotation is required as the mode is symmetric. All SEM image scale bars are 300 nm. The far-field emission pattern matches the simulation when considering a 120 degree rotation.

## Discussion

In conclusion, our work demonstrates that the scalable, deterministic nature of catalyst-free selective area epitaxy can be leveraged not only to achieve highly directional emission from InP nanowire pairs and triplets, but also to overcome limitations found in earlier demonstrations. Unlike prior methods requiring complex transfer or alignment processes, our approach directly grows vertically emitting nanowire lasers. By exploiting optical mode coupling, we lifted the degeneracy of the TE01 mode to generate distinct lasing modes with engineered directionality. This represents an improvement over most existing reports of nanowire lasers, where directionality is either uncontrolled or achieved through more intricate fabrication steps.

Our simulation-guided design confirmed the potential of this approach for tailoring emission directionality in nanowire lasing metasurfaces. Future investigations could focus on optimizing growth precision, exploring asymmetric nanowire arrangements, and realizing electrical carrier injection. Additionally, tuning the refractive index of the surrounding medium could potentially enable dynamic switching between lasing modes, further enhancing the versatility for integrated photonics and advanced optical beam-shaping applications.

# Materials and methods

## MOCVD growth

<111>A InP wafers were coated with approximately 200 nm $SiO_2$ using plasma enhanced chemical vapor deposition. Afterwards an EBL step was performed to create circular openings of approximately 100 nm diameter in a 200 nm thick mask of electron beam resist on top of the wafer. A subsequent inductively coupled plasma reactive ion etching step using $CHF_3$, created circular openings through the $SiO_2$ layer with a diameter of roughly 100 nm. Afterwards the remaining resist was carefully stripped by submersing the sample in acetone for approximately 30 minutes. Any further resist remains were stripped by low power oxygen plasma. Before growth, the sample was submerged for ~5 seconds in 1% diluted HF solution to remove any residual $SiO_2$ in the openings. A trim-etching step was performed with 30% $H_2O_2$ solution and 85% $H_3PO_4$, diluted with a ratio of 4 parts water to 1 part acid, to remove the top few monolayers on the InP substrate, which serves to repair any damage induced by the previous steps. First, the sample was dipped into the $H_2O_2$ solution for two minutes to oxidize the first layer, then cleaned with DI water, and then dipped into the $H_3PO_4$ solution for two minutes to remove the oxidized layer and then again cleaned with DI water. This was repeated 5 times.

Finally, the NWs were grown using metal organic chemical vapor deposition (MOCVD) (Aixtron 200/4 system), using $H_2$ as the carrier gas, trimethyl-indium and $PH_3$ as the precursors. Before the main growth step a 10-minute annealing step at 670 °C under $PH_3$ overpressure was performed. For the main growth step the temperature was increased to 680 °C with a V/III precursor ratio of 372. This step lasted 20 minutes.

## Optical characterization

A schematic illustration of the optical setup is shown in Figure S14. The excitation of the NWs was performed using the second harmonic of a solid-state Nd:YAG pulsed laser ($t_{pulse}$ = 7 ns, $f_{rep}$ = 100 Hz, $\lambda_{exc}$ = 532 nm). A 100x magnification, 0.9 NA Zeiss objective was used to focus down to a spot size of about 1 μm, which is sufficient to excite both NWs at the same time. Emitted PL emission was collected using the same objective. A grating (400 lines/mm, blazed at 500 nm) was used to spectrally resolve the emission (Princeton Instruments SP-2500i) and measured with a liquid nitrogen cooled front-illuminated CCD camera. All measurements were taken at room temperature.

The far-field patterns were taken by performing back-focal plane measurements. For this an additional lens was inserted into the PL setup after the objective, which served to collimate the PL emission based on emission direction rather than position. The rest of the setup remained identical. To combat the degradation of the NWs during measurements the exposure time was limited to 5 seconds. Far-field images were taken multiple times and added on top of each other to increase the signal to noise ratio. The patterns were measured just above the lasing threshold, where only one or a few Fabry-Perot modes dominate and again at well above the lasing threshold. Although at higher powers the patterns were clearer and more delineated from the PL background, the general shape of the patterns was unaffected. Shown are always the higher-power far-field patterns. The far-field images were background-corrected and smoothed slightly with a median filter to reduce noise.

## Simulations

Optical simulations were carried out using Lumerical. InP NWs were modeled as hexagons to elucidate general trends or as polygons to replicate pairs measured in experiments. Perfectly matched layers were used as the boundary condition. For the replication simulation of NW pairs the length difference was considered, but not the slightly pointy end-facet structure. The refractive indices of InP and $SiO_2$ was set to 3.4 [14] and 1.4, respectively, and the simulated wavelength was set to 850 nm, which corresponds to the central lasing wavelength of InP NWs. The extinction coefficient was set to 0, assuming transparency condition and to accurately calculate end facet reflectivities, independent of the lights travel distance beforehand. The NW system was injected with a specific mode by utilizing the mode source of Lumerical. A monitor was placed slightly above the NWs, which served to record the near-field emission. The far-field pattern was obtained through the in-built far-field projection functionality of Lumerical. Figure S15 illustrates schematically the two simulation setups used for the calculation of the reflectivities of the top and bottom NW end-facets respectively.

Array simulations were conducted by using the periodic boundary condition in x and y direction. The pitch in x and y direction was identical. Again, a mode source was used to inject light and a monitor was used to record the near field. 1, 3, 5, and 10 periods were used to calculate the far-field with a Gaussian weighting.

The gain-threshold calculation was done using both the MODE solver as well as the FDTD module of Lumerical. The effective and group index, as well as the confinement factor was calculated by the MODE solver. The end-facet reflectivity was calculated using FDTD simulations. To this end, a monitor was placed through the NW cross section which measured the back-reflected light from the end-facet. A mode decomposition of the obtained field-data was performed and the reflectivity determined as the fraction of input power to the reflected power in the original injected mode.

## Acknowledgements

We are grateful to Dr. Mykhaylo Lysevych for fruitful discussions and technical assistance. We acknowledge the Australian National Fabrication Facility (ACT node) for providing access to the epitaxial growth and fabrication facilities. This work was finally supported by the Deutsche Forschungsgemeinschaft (DFG, German Research Foundation) through the International Research Training Group (IRTG) 2675 "Meta-ACTIVE", project number 437527638 and the Australian Research Council through the Center of Excellence for Transformative Meta-Optical Systems.

## Author details

[1] Institute of Solid State Physics, Friedrich Schiller University Jena, Max-Wien-Platz 1, 07743 Jena, Germany

[2] ARC Centre of Excellence for Transformative Meta-Optical Systems, Department of Electronic Materials Engineering, Research School of Physics, The Australian National University, Canberra, ACT 2600, Australia

## Author contributions

All authors (L.R.J, W.W.W., C.R., H.H.T.) contributed to all aspects of the research with leading input from L.R.J.. H.H.T., C.R. and W.W.W. conceived and supervised the research. All authors (L.R.J, W.W.W., C.R., H.H.T.) contributed to writing the manuscript. L.R.J. performed the measurements, simulations, did the data analysis, prepared the figures and wrote the first draft of the paper.

## Data availability

Data underlying the results presented in this paper may be obtained from the authors upon request.

## Conflict of interest

The authors declare no competing interests

# Supplementary Information for

## Far-field directionality control of coupled InP nanowire lasers

*Lukas R. Jäger[1,2], Wei Wen Wong[2], Carsten Ronning[1], Hark Hoe Tan[2,*]*

[1] Institute of Solid State Physics, Friedrich Schiller University Jena, Max-Wien-Platz 1, 07743 Jena, Germany

[2] ARC Centre of Excellence for Transformative Meta-Optical Systems, Department of Electronic Materials Engineering, Research School of Physics, The Australian National University, Canberra, ACT 2600, Australia

* Corresponding author

## Section I: Calculation of the gain threshold $g_{th}$

The calculation of the gain threshold $g_{th}$, shown in Figure 1, is based on reference [15]. Each mode has a different gain threshold that depends on its confinement, end-facet reflectivity, and group index. It is given by:

$$g_{th} = \frac{1}{L\Gamma_0 n_g/n_b} \log\left(\frac{1}{r_1 r_2}\right) \quad \textbf{(S1)}$$

where $L$ is the NW length, $\Gamma_0$ is the physical electric field confinement in the NW, $n_g$ is the group index, $n_b$ is the background material index (3.4), and $r_1$ and $r_2$ are the end-facet reflectivities. The end-facet reflectivities were calculated using FDTD simulations. The confinement factor $\Gamma_0$ is calculated from

$$\Gamma_0 = \frac{\int_{cavity} dA\, w_e}{\int_{waveguide} dA\, w_e} \quad \textbf{(S2)}$$

where $w_e$ is the stored electric field energy.

## Section II: Electric mode profiles of unequal NWs in a pair structure

We calculated the electric mode profiles of NW pairs with different cross-section pairings to corroborate the claim that these profiles remain largely unaffected by fabrication inaccuracies. In particular, the fabricated NWs often exhibited a polygonal cross-section, such as a dodecagon. We simulated hexagon–hexagon, hexagon–circle and circle–circle cross-sections pairings, considering a circle as an extreme case: a polygon with infinite sides.

Figure S1 shows the resulting electric field distributions and polarization directions of the TE01 modes for these pairings. The doughnut-like shape of the TE01 mode persists when transitioning from hexagonal to circular cross-sections, as well as in hybrid pairings. In the asymmetric pairing of circular and hexagonal cross-sections, the TE01-a mode exhibits stronger electric field intensity in the circular NW, whereas in the TE01-b mode, the hexagonal NW carries more power. In the equal pairings both NWs carry the same power for both modes. This asymmetry is due to the difference in cross-sectional area between the NWs in the pair. The same effect was observed when solving the mode profiles for NW pairs with equal cross-section shapes, but with different NW diameters, and therefore areas.

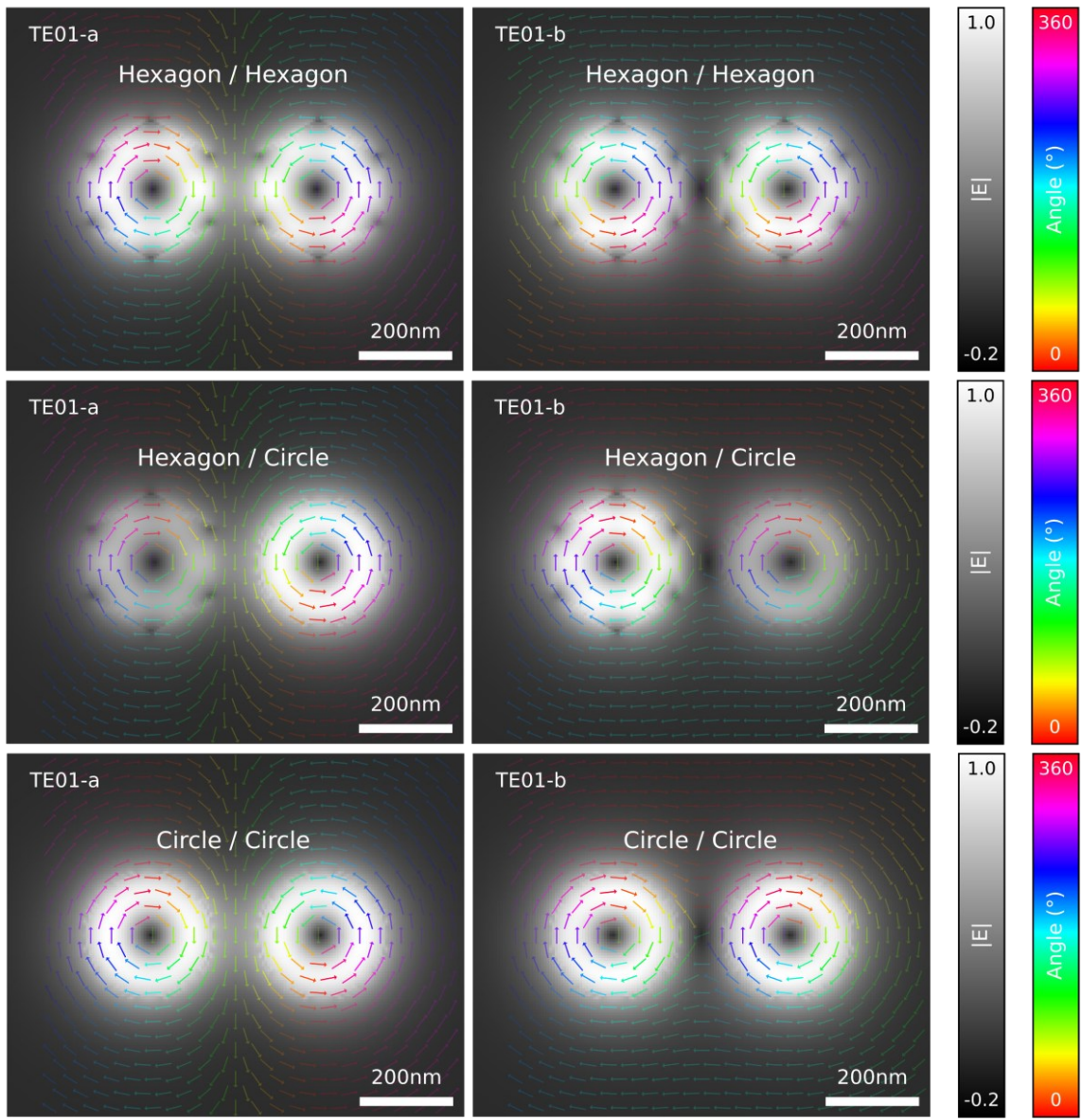

**Figure S1:** Normalized electric field strength of simulated mode profiles in the NW pair structure. The NWs have different shapes as indicated, and a gap of 100 nm. The arrows denote the normalized polarization direction of the electric of magnetic field. The color of the arrows encodes the angle.

## Section III: Electric mode profiles of a NW pair

Figure S2 shows the electric field strength and polarization of the first eight modes in the NW pair system. Each mode of an isolated NW splits into two when coupled in a pair, lifting the polarization degeneracy inherent in the single-NW case. Therefore, the two resulting modes exhibit field profiles that are either in phase or out of phase across the NWs. For the HE11 mode family, the in-phase modes, and for the TE01 and TM01 families, the out-of-phase modes have the higher effective index.

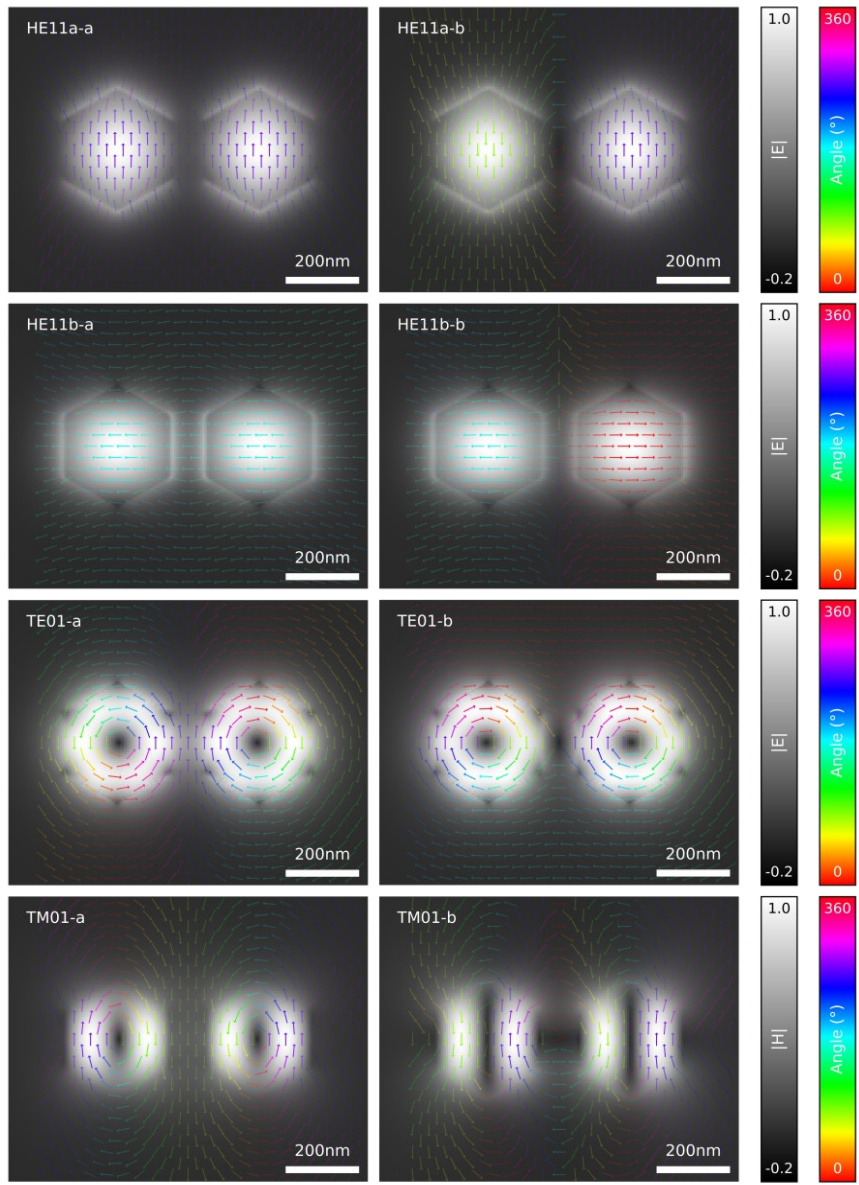

**Figure S2:** Normalized electric or magnetic field strength of simulated mode profiles of a pair of NWs with a diameter of 340 nm and a gap of 100 nm. The arrows denote the normalized polarization direction of the electric of magnetic field. The color of the arrows encodes the angle.

## Section IV: Origin of the NW pair far-field emission directionality

The TE01-a and TE01-b modes of the NW pair, which arise from the single NW TE01 mode, exhibit distinct and orthogonal far-field emission patterns. Similar patterns are observed in simulations of two closely spaced dipole sources, as shown in Figure S3. The far-field emission patterns of the dipole pair resemble those of the TE01-a and -b modes when the dipoles are 180° out of phase and in phase, respectively. The electric field patterns, described in SI Section III, support the interpretation of the NW end facets as approximating in-phase and out-of-phase dipole pairs. Minor deviations between the emission patterns of the NW and the dipole systems likely arise from differences in the emission characteristics of their constituent emitters.

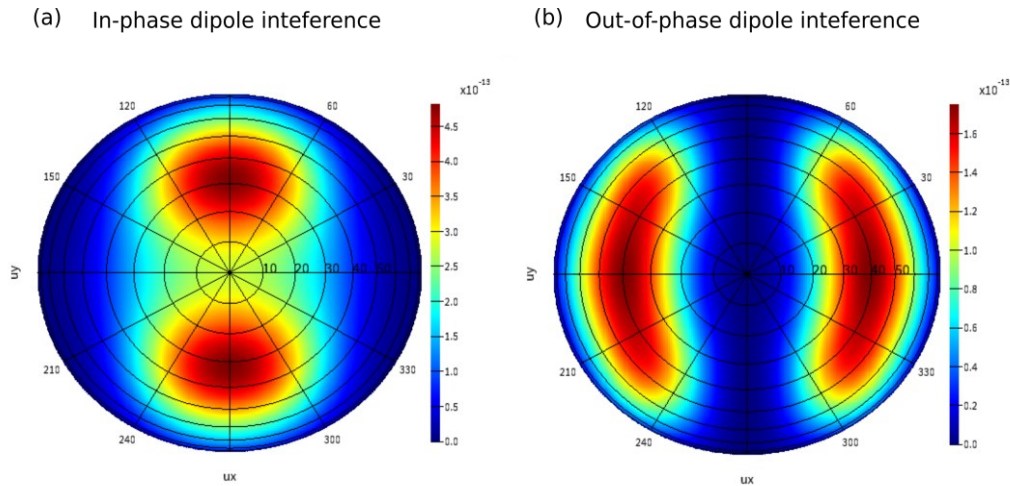

**Figure S3:** Far-field simulations of a dipole pair. The dipoles are separated by 400 nm with 850 nm emission wavelength and direction as indicated in the figure. **(a)** No phase difference of the dipoles. **(b)** 180° phase difference.

## Section V: Below threshold NW pair PL measurement

The detailed features of the lowest pump fluence PL spectrum shown in Figure 2a are obscured by the higher intensity of the spectra at higher pump fluences. To highlight the PL spectrum below threshold, it is replotted in Figure S4. There is a clear NBE peak at approximately 875 nm.

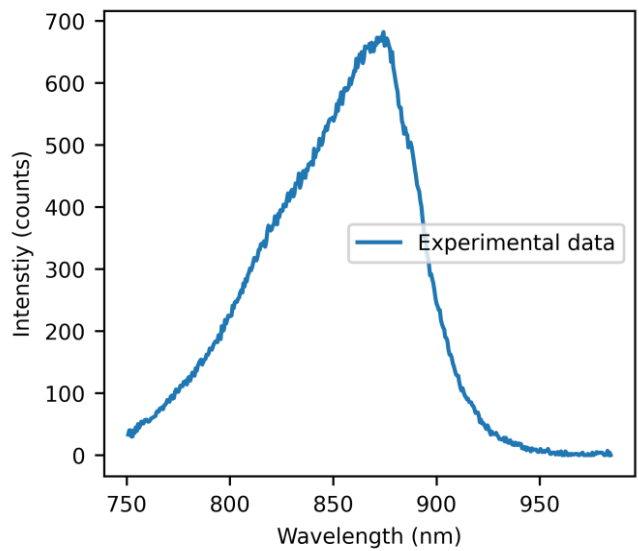

**Figure S4:** PL data of a NW pair at low pump fluence (85 µJ cm$^{-2}$ pulse$^{-1}$) from figure 2a in the main text.

# Section VI: Modeling the gain of WZ-phase InP

Fitting of the light-in–light-out (LL) curve in Figure 2c requires substantial background information. First, we describe how the gain of bulk WZ-phase InP is calculated. This gain is then used to solve the laser rate equations, based on the "ABC" model, which incorporates nonradiative, radiative and Auger recombination processes. This approach closely follows the methodology outlined in reference [16]. The parameters used in the modeling are summarized in Table S1.

## Calculating the material gain

The material gain is calculated using the following expression [14]:

$$g(\hbar\omega) = \frac{\pi e^2}{nc\epsilon_0 m_0^2 \omega} |M|^2 \int \rho_r(E) * (f_c(E) - f_v(E)) * l(E - \hbar\omega) dE \tag{S3}$$

where $e$ is the electron charge, $n$ is the bulk material refractive index, $c$ is the vacuum speed of light, $\epsilon_0$ is the vacuum permittivity, $m_0$ is the electron rest mass, $\hbar$ is the reduced Planck constant, $\omega$ is the angular frequency of the photon, $|M|^2$ is the momentum matrix element of bulk InP, $\rho_r(E)$ is the reduced density of states (DOS), $f_c(E)$ and $f_v(E)$ are the Fermi-Dirac carrier distribution functions of the conduction and valence band, respectively, and $l(E - \hbar\omega)$ is the homogeneous line shape broadening function.

The reduced DOS $\rho_r$ is given by

$$\rho_r = \frac{1}{2\pi} \left(\frac{2m_r}{\hbar^2}\right)^{3/2} E^{\frac{1}{2}} \tag{S4}$$

where the reduced effective carrier mass $m_r$ is given by

$$m_r = \frac{m_e^* m_h^*}{m_e^* + m_h^*} \tag{S5}$$

The linewidth describing line broadening is given by [7]

$$l(E - \hbar\omega) = \frac{1}{\pi\gamma} \operatorname{sech}\left(\frac{E - \hbar\omega}{\gamma}\right) \tag{S6}$$

where $\gamma$ is the Lorentzian or homogeneous line width, and a fitting parameter.

To calculate the quasi-Fermi levels, we utilize the charge neutrality equations, which are given by

$$N = N_c F_{\frac{1}{2}}\left(\frac{F_c - E_c}{k_B T}\right) = 2\left(\frac{m_e^* k_B T}{2\pi\hbar^2}\right)^{\frac{3}{2}} \tag{S7}$$

$$N = N_v F_{\frac{1}{2}}\left(\frac{E_v - F_v}{k_B T}\right) = 2\left(\frac{m_h^* k_B T}{2\pi\hbar^2}\right)^{\frac{3}{2}} \tag{S8}$$

To solve the Fermi-Dirac integral of order ½, we employ a numerical approximation [17], which enables us the use of a numerical root-solving algorithm. Solving the charge neutrality equations yields the quasi-Fermi levels $F_c$ and $F_v$, which are used to compute $f_c(E)$ and $f_v(E)$ [18]. The temperature-dependent bandgap $E_G(T)$ is computed using a modified Varshni equation for WZ InP [17]

$$E_G(T) = E_B - a_{WZ}[1 + 2/(e^{\frac{\theta}{T}} - 1)] \tag{S9}$$

The gain is computed for a finely sampled series of 64 charge carrier density samples, ranging from $10^{15}$ cm$^{-3}$ to $10^{21}$ cm$^{-3}$. Intermediate values are estimated via interpolation. Figure S5a displays the calculated gain as a function of carrier density for lasing wavelengths of 800, 850 and 900 nm, demonstrating that the gain saturates with increasing carrier density. Figure S5b depicts the gain profiles at various carrier densities for a fixed wavelength of 850 nm. As the carrier density increases, the gain envelope broadens towards shorter wavelengths, due to the greater availability of states in the valence and conduction bands at higher excitation levels.

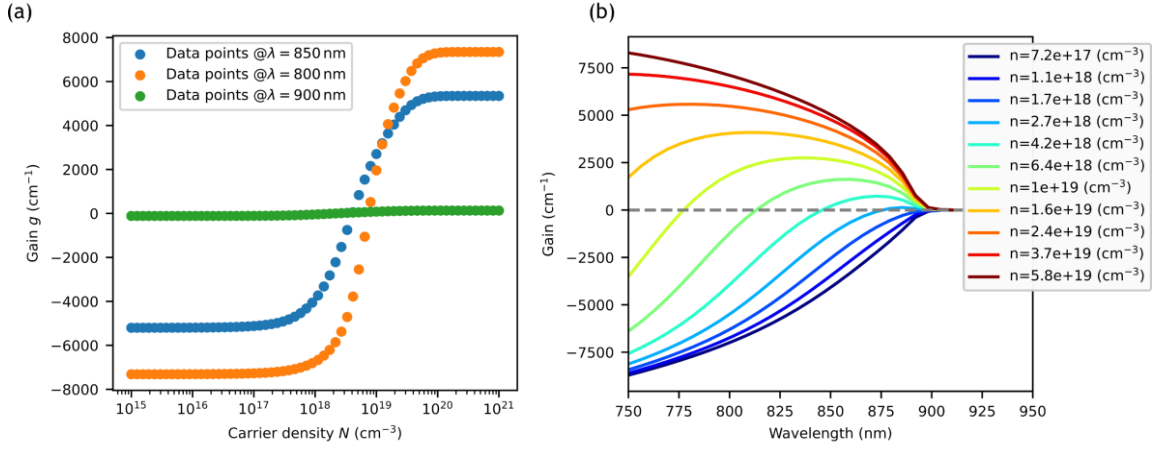

**Figure S5:** Results of calculating the gain of bulk WZ-phase InP. **(a)** Calculated gain for a select number of wavelengths vs free carrier density. **(b)** Calculated gain vs wavelength for various free carrier concentrations, $n$.

## Estimation of the injection efficiency

The pump efficiency is calculated as:

$$\eta_{pump} = \frac{\sigma_{abs}}{A_{spot}} \approx 40.7\% \tag{S10}$$

The absorption cross section, $\sigma_{abs}$, is a measure for the probability of photon absorption within the material. It was calculated using an FDTD simulation with the Adachi n-k values for InP in the 800 to 900 nm range [19]. The NW volume $V$ is calculated as:

$$V = N_{NW} L_{NW} A_{shape} \approx 0.23\ \mu m^3 \tag{S11}$$

$$A_{shape} = A_{hexagon} = \frac{3\sqrt{3}}{2}\left(\frac{D}{2}\right)^2 \tag{S12}$$

where $N_{NW}$ is the number of NWs, $L_{NW}$ is the NW length, $A_{shape}$ is the area of the NW cross section, and $D$ is the NW diameter.

## Laser rate equation

The standard "ABC" model [18] is used to describe the carrier and photon dynamics in the laser system. The coefficients $B$ and $C$, corresponding to radiative and Auger recombination, are taken from literature values [20]. The coefficient $A$ is estimated from the inverse of the minority carrier lifetime ($A = \frac{1}{\tau_{nr}} \approx \frac{1}{\tau_{mc}}$), where $\tau_{nr}$ and $\tau_{mc}$ denote the non-radiative and minority carrier lifetimes, respectively. The rate of change in carrier density $N$ is given by [18]:

$$\frac{dN}{dt} = \frac{\eta_{pump} P(t)}{E_{pump} V} - AN - BN^2 - CN^3 - v_g g(N) S \tag{S13}$$

while the change in photon number S is given by:

$$\frac{dS}{dt} = \Gamma(v_g(g(N) - g_{th})S + \beta BN^2) \tag{S14}$$

where the pump energy $E_{pump}$ is given by $E_{pump} = \hbar\omega$ and the group velocity $v_g$ is calculated using the group index $n_g$ by $c_0 = \frac{v_g}{n_g}$. The group index $n_g$ is estimated from Lumerical MODE simulations to be approximately 4.5. However, these simulations are performed without material dispersion. To obtain a more realistic estimate of the group index, we analyze the mode-spacing in a measured spectrum, assuming the observed modes are Fabry–Pérot resonances, as depicted in Figure S6. This yields a large value for $n_g \approx 11.4$. For the subsequent rate equation modeling, we use $n_g = 10$. The confinement factor is estimated to be $\Gamma \approx 0.95$.

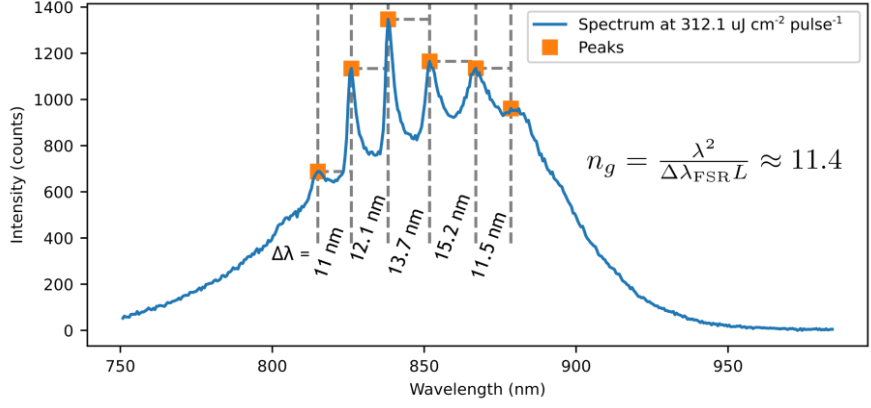

**Figure S6:** One of the spectra shown in figure 2. Indicated are the peak maxima and the wavelength spacing between them. Assuming that they are Fabry-Perot modes the group index $n_g$ can be estimated from the free free spectral range $\Delta\lambda_{FSR}$ and NW length $L$ ($\approx 5\ \mu m$).

The temporal profile of the pump laser is modeled as [7]
$$P(t) = P_{peak} \operatorname{sech}^2(t/\tau) \tag{S15}$$

$$P_{peak} \approx 0.88 \frac{E_{pulse}}{\Delta t_{FWHM}} \tag{S16}$$

where $P_{peak}$ is the peak power of the laser pulse and $\Delta t = 400$ fs is the temporal pulse width.

The rate equations are integrated using the Radau method [21] to improve computational efficiency. At low pump powers the photon number decays slowly, requiring extended integration times before the integration could be stopped. At high pump powers, the integration is terminated once the photon number falls below 0.01% of the peak value. The maximum integration time is set to 10 ns. In Figure S7a the carrier and photon densities are shown on two timescales to illustrate the fall-off speed. Figure S7b shows the calculated LL-curve for three values of $\beta$, demonstrating the characteristic S-shaped response.

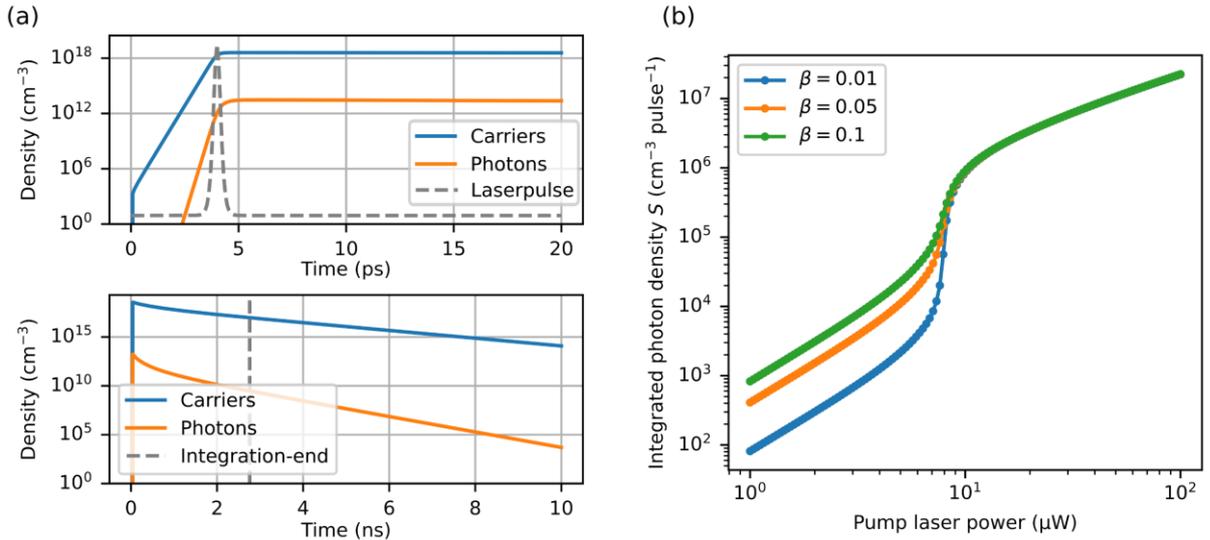

**Figure S7:** Solving the laser rate equations for a gain threshold of 1000 cm$^{-1}$. **(a)** The densities of carriers and photons over time, at long and short timescales. The laser pump power was 5.5 µW and $\boldsymbol{\beta = 0.01}$. The dashed line labeled "Integration-end" indicates until what time the differential equations were solved. The stopping point was determined when the photon density fell below 0.01% of the maximum. The data points after the integration-end are a result of extrapolation. The maximum integration time was 10 ns. **(b)** Calculated LL-curve for different $\boldsymbol{\beta}$-values.

To fit this function to the experimental data, we first calculate the LL curve, as shown in Figure S7b, for given values of $\beta$ and the gain threshold. The resulting integrated photon number is then scaled to align with the measurements. An additional background parameter is included to account for any stray light in the measurement setup. Due to the computational cost of the calculation of a LL curve, automated fitting procedures are impractical. Instead, the parameters are manually adjusted to achieve a decent fit.

**Table S1** Material parameters, corresponding meaning and values used in the calculation of InP WZ-phase gain

| Material parameter | Parameter meaning | Value |
|---|---|---|
| n | Refractive index | 3.4 [14] |
| $E_p(\|M\|^2 = \frac{m_0 E_p}{6})$ | | 20.7 eV [14] |
| $E_B$ | | 1.545 eV [22] |
| a | | 45 meV [22] |
| θ | Modified Varshni equation parameter | 238 K [22] |
| $m^*_e$ | Effective electron mass | 0.088 $m_0$ [22, 23] |
| $m^*_h$ | Effective hole mass | 0.37 $m_0$ [22, 23] |
| $v_g$ | Group velocity | 10 |
| Γ | Confinement factor | 0.95 |
| A | Non-radiative recombination coefficient | $\approx \frac{1}{\tau} = 9.09 * 10^8 s^{-1}$ [16] |
| B | Radiative recombination coefficient | $6.25 * 10^{-10}\ cm^3 s^{-1}$ [20]] |
| C | Auger recombination coefficient | $9.1 * 10^{-31} cm^6 s^{-1}$ [20] |
| γ | Linewidth broadening | 0.005 [16] |

# Section VII: Origin and simulation of the far-field ripple pattern

The ripple or interference pattern observed in Figures 3 and 5 arises from the interference of light scattered from both end-facets, as schematically depicted in Figure S8a. Two pieces of evidence support this interpretation. First, we conducted a 2D simulation of an entire single NW system. Capturing interference at large emission angles requires the simulation region to be significantly wider than it is tall. A typical NW is approximately 5 μm long, which would make a full 3D simulation computationally prohibitive in terms of both runtime and memory usage. The simulated NW length and the simulation region width were set to 4 μm and 40 μm, respectively. This 10:1 aspect ratio allows the capture of far-field interference up to approximately 84°. Two mode sources were used, each injecting the fundamental TE mode—the lowest-order mode with a central minimum, equivalent to the TE01 mode in 3D simulations. One source was directed upward and the other downward. This dual-source configuration was chosen to ensure a balanced scattering contribution from both end-facets. As shown in Figure S8b, the resulting far-field intensity profile exhibits a clear interference pattern. The interference pattern is weaker at the center and stronger further out. Close to the maximum resolvable angle (84°), the interference pattern diminishes due to the limited aspect ratio of the simulation region.

Further insight is obtained by comparing the far-field emission patterns and interference ring spacings between two NW systems of differing lengths. Figure S9 displays the far-field pattern from Figure 3 alongside tat of a single, longer NW. The pattern from the longer NW exhibits a greater number of fringes with smaller spacing. This behavior is consistent with the interpretation that the pattern arises from interference between light scattered at the NW end-facets. The lengths of the pair and single NW are 5 μm and 6.7 μm, respectively.

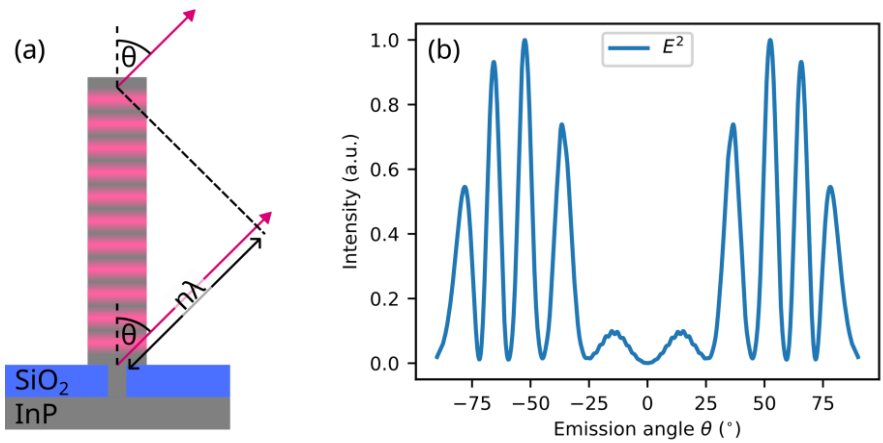

**Figure S8:** Explanation of the interference fringes in the far-field emission pattern **(a)** Schematic side view of a single NW. Scattered light from the top and bottom end-facets of the NW interferes in the far-field if the path length difference is a multiple of the wavelength. **(b)** Simulated far-field data of a two-dimensional system. A NW of length 4 µm and 260 nm diameter was simulated on 40 µm wide grid, to capture large angle interference. (A two-dimensional simulation was used as this is unfeasible in a three-dimensional simulation due to memory constraints).

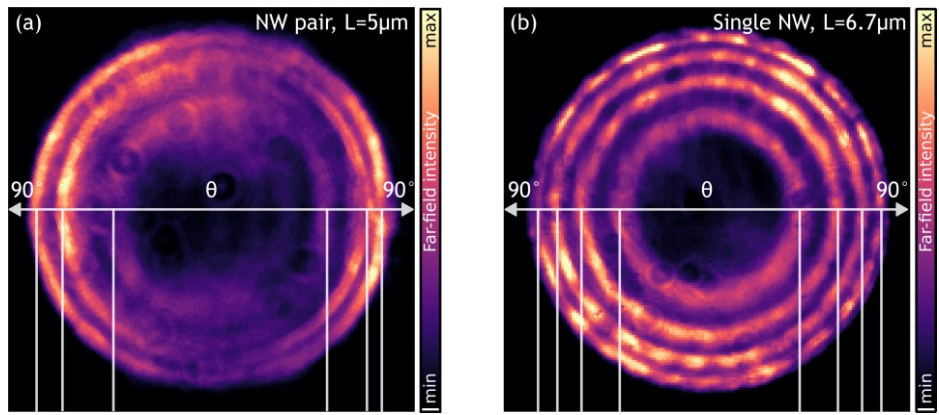

**Figure S9:** Comparison of the interference pattern of **(a)** the NW pair from figure 3 and **(b)** a single NW. The single NW is significantly longer (6.7 µm compared to the 5 µm long pair). The single NW far-field pattern has more and closer spaced interference rings. This is consistent with the explanation that the ripple pattern stems from the interference of the two NW end-facets.

## Section VIII: Farfield patterns of the TE01-a and TE01-b modes at various gaps

Figure S10 shows simulated far-field patterns of NW pairs lasing in the TE01-a (Figure S10a) and TE0-b (Figure S10b) modes. At a gap of 10 nm, the far-field pattern corresponding to the TE01-a mode shows a normal emission lobe. The intensity of the lobe diminishes at higher gaps, while the large solid angle emission lobes gain intensity. The far-field patterns of the TE01-b mode do not show a normal lobe at small gaps, but instead, two large angle emission lobes, which exhibit a diminishing width with increasing gap. At a gap of approximately 800 nm, both patterns exhibit interference stripes.

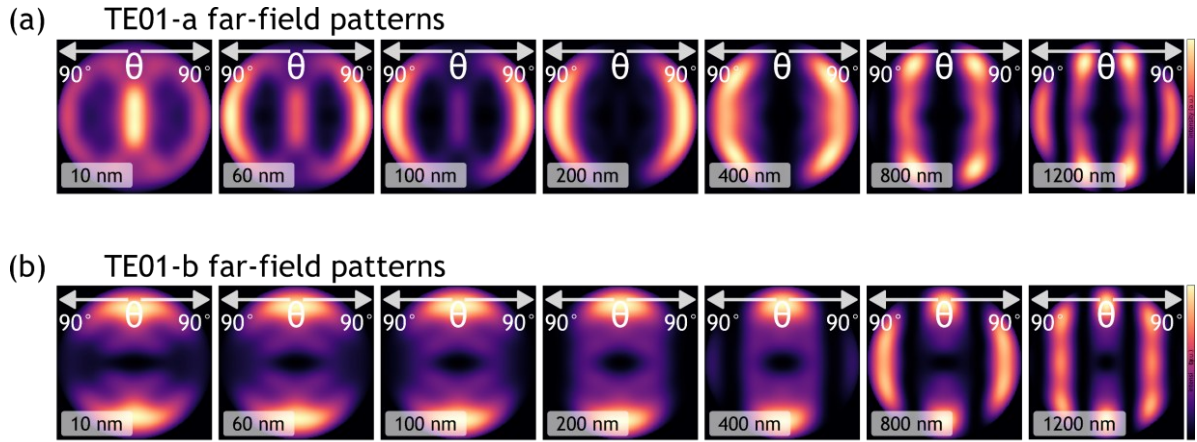

**Figure S10:** Far-field patterns of a NW pair with 250 nm diameter for a range of gaps as indicated. **(a)** TE01-a mode **(b)** TE01-b mode.

## Section IX: TE01-b mode far-field emission pattern

When two NWs are coupled, the TE01 mode of a single NW splits into two distinct modes. NW pairs lasing in the TE01-b mode emit directional light, similar to the TE01-a mode, but with the key difference that the primary emission direction is orthogonal. In simulations assuming ideal geometry (perfect hexagonal cross-sections and equal diameters and lengths) the TE01-a mode exhibits a lower gain threshold. However, fabricated NWs often deviate from the ideal geometry due to growth imperfections. As a result, some NW pairs lase preferentially in the TE01-b mode, producing far-fields that differ significantly from those expected in the designed TE01-a mode. Figure S11a shows the measured far-field pattern of a representative NW pair lasing in the TE01-b mode, while Figure S11b shows the corresponding simulation. A SEM image of the pair (Figure S11c) reveals that one of the NWs has a significantly smaller diameter and a slightly shorter length.

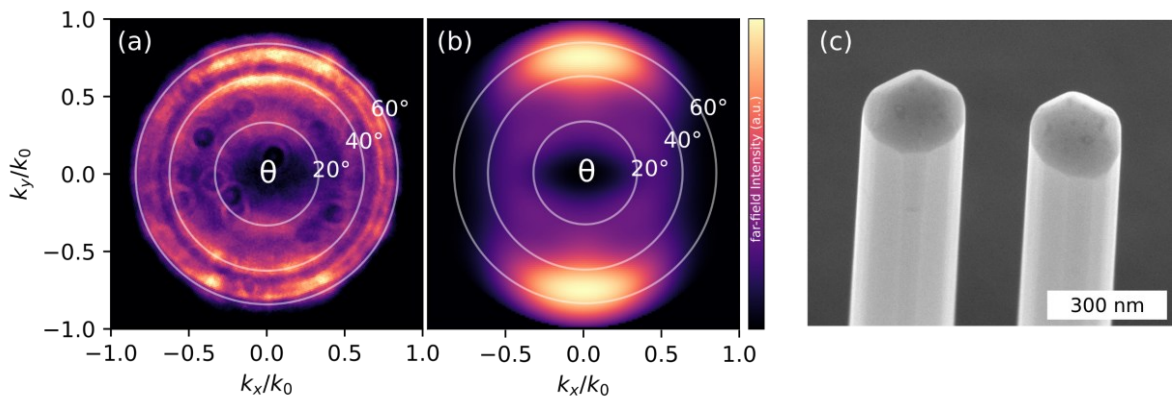

**Figure S11:** TE01-b mode **(a)** experimental back-focal plane data, **(b)** simulated far-field emission, (c) SEM image of the InP NW pair measured in (a). Note that the emission direction is orthogonal to the one of the TE01-a mode.

## Section X: Array of NW pairs mode profiles and dipole approximation

Figure S12a shows the injected electric mode profiles used to generate the far-field patterns depicted in Figure 4b. The unit cell mode profiles closely match the TE01-a mode of a single NW pair.

Figure S12b shows the simulated far-field patterns produced by an array of sub-wavelength-spaced dipole pairs, analogous to the NW pairs. The panels from left to right correspond to decreasing array pitch. Dipoles were spaced 400 nm apart, with a simulation wavelength of 850 nm. Far-field calculations were performed over three periods using Gaussian weighting. All simulated patterns closely resemble those of the NW pair array system. In particular, the array with a 1000 nm pitch exhibits two narrow emission lobes. For the 800 nm pitch, the dipole array produces two weak, near-central lobes—slightly different from the NW system, which exhibited a single central lobe.

Deviations between patterns of the same pitch can be attributed to the differing emission characteristics of the underlying unit emitters: individual NWs exhibit cone-like emission, whereas dipole sources exhibit a dipolar radiation pattern.

These findings suggest that the NW pair array exhibits narrower emission lobes as a result of interference.

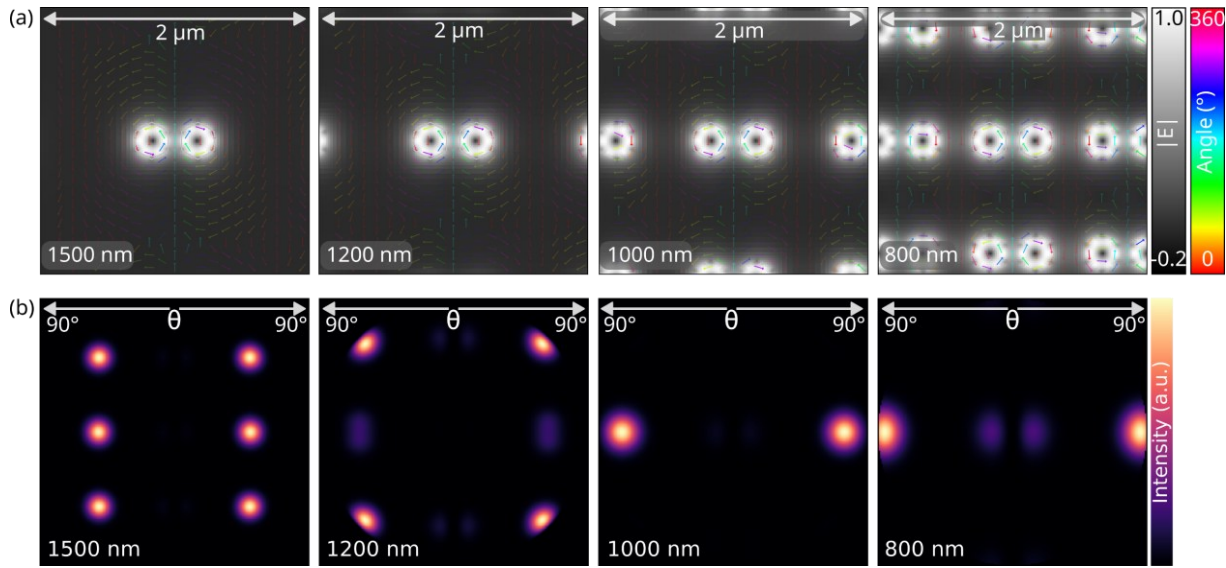

**Figure S12:** **(a)** Electric mode profiles injected into an array of NW pairs, corresponding to the produced far-fields depicted in Figure 4b. The pitch is noted in the lower left corner. **(b)** Far-fields produced by analogous simulations using a dipole source pair as the unit cell.

## Section XI: Electric mode profiles of the NW triplet system

Figure S12 displays the electric field strength and polarization vectors of the three dominant modes in the NW triplet system, which originate from the TE01 mode of a single NW. The modes are labeled based not on their effective indices, but on the number of NWs in which the majority of the optical power is confined. The TE01-2 mode has the highest effective index, followed by TE01-1 and then TE01-3. In addition, the fundamental HE11a and HE11b modes of the single NW split into six distinct modes in the triplet system (data not shown).

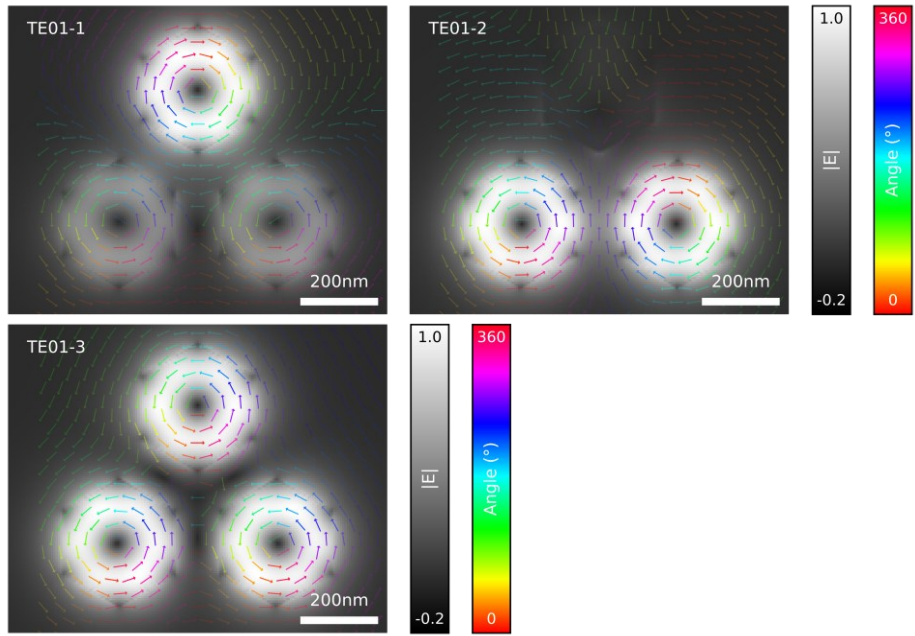

**Figure S13:** Normalized electric field strength of simulated mode profiles in the triplet NW structure. The NWs have a diameter of 320 nm and a gap of 120 nm. The arrows denote the normalized polarization direction of the electric of magnetic field. The color of the arrows encodes the angle.

## Section XII: PL measurement of a grown InP NW pair array

Although the fabricated InP NW pair array exhibited significant deviations from the intended geometry, specifically, an average NW diameter of ~100 nm, well below the target of 250 nm, we conducted PL measurements using an enlarged laser spot to simultaneously excite multiple unit cells. Figure S13 shows PL measurement from an array with pitch sizes 800 nm (x-direction) and 600 nm (y-direction). The diameter of the laser spot was about 20 μm, and therefore covered a large number of unit cells. The PL signal, as shown in Figure S13b, increases with increasing pump fluence. At low pump fluences, a broad peak near 860 nm stemming from the WZ-phase NW NBE emission is visible. At higher pump fluences, a second peak emerges at about 815 nm, which is attributed to a collective resonant mode of the pair array. However, this peak does not become dominant in the spectrum, as would be expected for a lasing emission peak. At high pump fluences, the substrate ZB-phase InP NBE emission peak becomes visible. PL measurements were conducted at the Australian National University using a femtosecond-pulsed femtoTRAIN IC-Yb-2000 laser, frequency-doubled to 522 nm, with a pulse duration of 400 fs and a repetition rate of 10.2MHz. A 60×, 0.7 NA aberration corrected objective (Nikon CFI Plan Flour) was used to focus the laser beam. A spherical lens was used to diverge the laser beam, thereby enlarging the laser spot size on the sample. The emission was collected through the same objective, filtered to remove the reflected pump light, and spectrally resolved grating spectrometer (Acton SpectraPro 2750). The same setup, using a focused laser spot of 1 μm, was employed to record the data for the LL curve shown in Figure 2c.

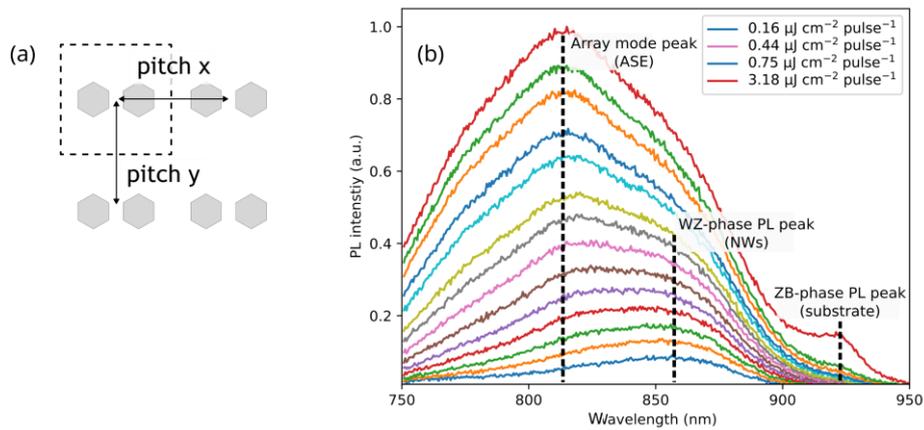

**Figure S14:** **(a)** Schematic top-down view of the array structure. **(b)** PL spectra of an array of InP NW pairs at different excitation powers. The array was grown with an x-pitch, y-pitch and NW center-to-center distance of 800, 600, and 340 nm, respectively. Three peaks are indicated by dashed lines: Array mode peak (ASE), WZ-phase PL peak (NWs), and ZB-phase PL peak (substrate).

## Section XIII: Optical measurements setup

The optical setup used to acquire the back-focal plane images is depicted in Figure S14. The NWs were excited using a frequency-doubled Nd:YAG laser (InnoLas SpitLight Compact) operating at 532 nm, with 7 ns pulse duration and a 100 Hz repetition rate. Laser power was monitored using a photodiode (Thorlabs S130VC). A 100×, 0.9 NA Zeiss objective was used both to focus the excitation beam and to collect the emitted light. A spherical lens was positioned after the objective at a distance of 2f from the back-focal plane to image it. The imaged back-focal plane was captured by a nitrogen cooled CCD.

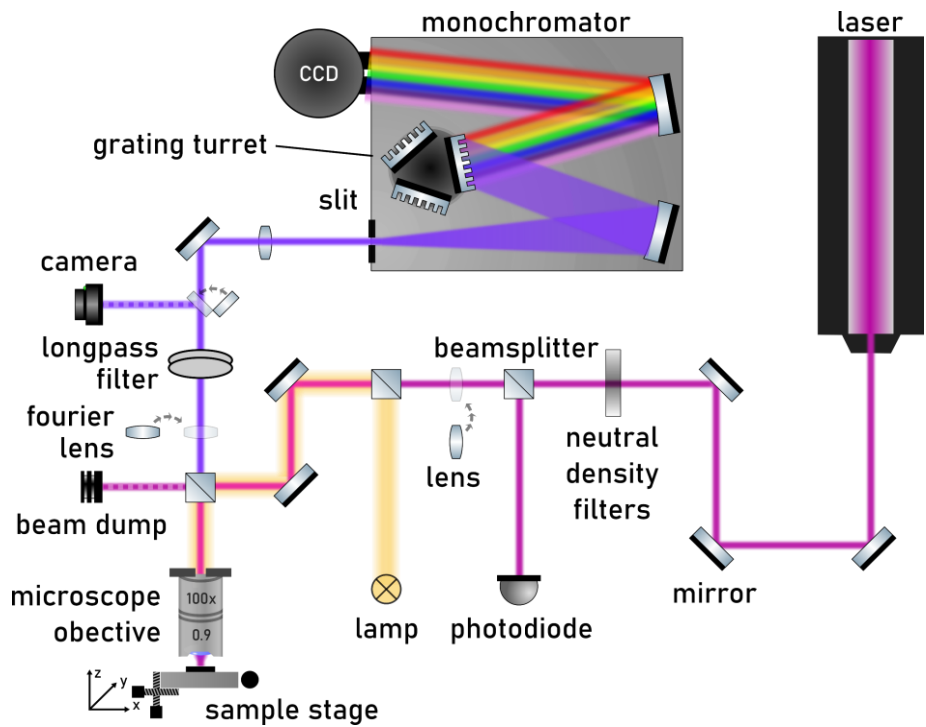

**Figure S15:** Schematic of the PL and BFP measurement setup.

## Section XIV: Simulation setup

All simulations were performed using the MODE and FDTD solver in Ansys Lumerical. For 3D FDTD simulations, the "auto non-uniform" mesh setting was employed an accuracy level of 4. Convergence testing confirmed that this setting provided sufficient accuracy. In the MODE simulations, a mesh refinement was used to increase the resolution to 6 nm within the NW region and approximately 50 nm beyond it.

To reduce simulation time and memory usage, only the final micrometer of the NW system was simulated. This is illustrated in Figure S15. In both simulation cases—the top and bottom end-facets of the NW system—a power monitor was placed close to the mode source to measure reflected light. For top end-facet simulation, as shown in Figure S15a, a nearfield power monitor was positioned 100 nm above the end-facet. The monitor data was post-processed using Lumerical's built-in far-field projection function to compute the electric field distribution on a hemispherical surface at a distance of 10 m. The bottom end-facet simulation, shown in Figure S15b, did not require a second power monitor.

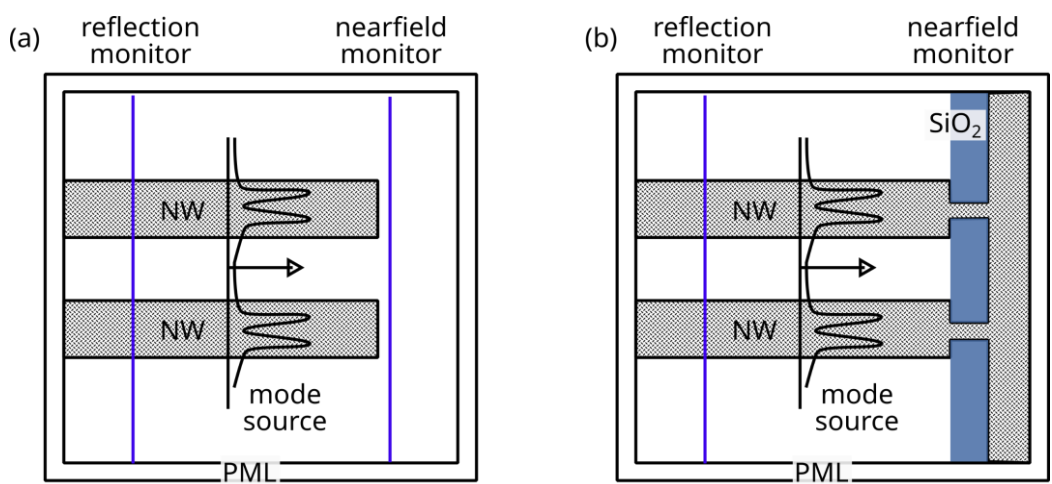

**Figure S16:** Simulation setup of **(a)** the top end-facet and **(b)** the bottom end-facet. The simulation setup of a single NW or a NW triplet is equivalent.